\documentclass[a4paper]{article}
\usepackage{amsmath}
\usepackage{amsfonts}
\usepackage{amsthm,amssymb,upref,amscd}  
\usepackage[T1]{fontenc}
\usepackage[english]{babel}
\usepackage[latin1]{inputenc}
\usepackage{ae}
\usepackage{aecompl}
\newtheorem{theorem}{Theorem}
\newtheorem{lemma}[theorem]{Lemma}

\newtheorem{definition}[theorem]{Definition}

\title{Axisymmetric stationary solutions with arbitrary multipole moments}
\author{Thomas B\"ackdahl\thanks{Department of  
Mathematics, Link{\"o}ping University,
SE-581 83 Link{\"oping}, Sweden.\newline
\hspace*{5mm} e-mail: thbac@mai.liu.se}}
\date{}

\begin{document}
\maketitle
\begin{abstract}
In this paper, the problem of finding an axisymmetric stationary spacetime from 
a specified set of multipole moments, is studied. The condition on the multipole moments,
for existence of a solution, is formulated as a convergence condition on 
a power series formed from the multipole moments. The methods in this paper can 
also be used to give approximate solutions to any order as well as estimates 
on each term of the resulting power series.
\end{abstract}
\section{Introduction}
The concept of multipole moments for curved spacetimes has been around for more than 35 years,
but there are still some unanswered questions.
A geometrical definition of multipole moments for static asymptotically flat spacetimes
was given by Geroch \cite{geroch}, and was generalised to the stationary case by Hansen \cite{hansen}.
If one also incorporates Beig's \cite{beigAPA} generalisation of centre of mass, one get a completely
coordinate independent definition of the multipole moments.

One of the remaining questions to answer is whether, for any given set of multipole moments,
there is a solution to Einstein's vacuum field equations with those multipole moments.
There are solution generating techniques, for instance, for the axisymmetric case, the FHP-method \cite{FHP}
can be used. However whether these methods always converge is still an open question.
In the general static case recent work has been done by Friedrich \cite{friedrich}. However, he establishes
results for entities related to the multipole moments, but not for the multipole moments themselves.
In the static axisymmetric case the complete convergence conditions were established in \cite{backdahl1}.
The convergence problem for the stationary axisymmetric case is treated in this paper.

The tensorial recursion of Geroch and Hansen can be simplified to a power-series expansion of a function 
formed from the metric components. This was shown in \cite{backdahl2} for the stationary axisymmetric case
and in \cite{backdahl3} for the general stationary case. 
It is easy to see \cite{backdahl3} that a necessary condition on the multipole moments, for existence of 
a solution, is that this power-series expression has positive radius of convergence. 
In this paper, this condition is also proven to be sufficient for the axisymmetric case.

\section{Field equations and potentials}
In this paper we restrict our selves to stationary vacuum spacetimes. 
In this section and the next we will treat the general stationary case but will later 
restrict our selves to the axisymmetric case.

Consider a stationary spacetime $(M, g_{ab})$ with time-like Killing vector field $\xi^a$.
Let $V$ be the 3-manifold of trajectories of $\xi^a$, i.e., $M$ with time factored out.
We let $\lambda=-\xi^a\xi_a$ be the norm of $\xi^a$.
Furthermore, we define the twist $\omega_a$ through 
$\omega_a=\epsilon_{abcd}\xi^b\nabla^c\xi^d$.
Due to the fact that we only consider vacuum solutions, we can define a twist potential
\footnote{Defined in a region who's spatial part is $\mathbb{R}^3$ minus a ball.}
$\omega$ via $\nabla_a\omega=\omega_a$, since $\nabla_{[b}\omega_{a]}=0$ in vacuum.

The metric $g_{ab}$ (with signature $(-,+,+,+)$) on the spacetime induces the positive definite metric
$$h_{ab}=\lambda g_{ab}+\xi_a\xi_b$$ on $V$.

In the stationary vacuum case, Einstein's field equations induce field equations on $V$ (rewritten from \cite{geroch71}),
\begin{subequations}
\begin{align}
R_{ab}&=\frac{1}{2\lambda^2}((D_a\lambda)D_b\lambda+(D_a\omega)D_b\omega)\\
D^aD_a\lambda&=\frac{1}{\lambda}((D^a\lambda)D_a\lambda-(D^a\omega)D_a\omega)\\
D^aD_a\omega&=\frac{2}{\lambda}(D^a\lambda)D_a\omega,
\end{align}
\end{subequations}
where $R_{ab}$ and $D_a$ is the Ricci tensor and the covariant derivative operator with respect 
to the metric $h_{ab}$.
We introduce the potential 
\begin{equation}\label{phidef}
\phi=\frac{1-\lambda-i\omega}{1+\lambda+i\omega} .
\end{equation}
With this potential, the field equations reduce to
\begin{subequations}
\begin{align}\label{riccieq}
R_{ab}&=\frac{2}{(\phi\bar\phi-1)^2}(D_{(a}\phi)D_{b)}\bar\phi\\
\label{ernst}
D^aD_a\phi&=\frac{2\bar\phi}{\phi\bar\phi-1}(D^a\phi)D_a\phi.
\end{align}
\end{subequations}
The equation \eqref{ernst} is a useful form of the Ernst equation \cite{ernst}. 
The non-linearity of this equation makes it quite difficult to solve
in general. In this paper however we will prove existence of solutions when
the multipole moments are specified.

\section{Definition of multipole moments}
In this section, the definition of multipole moments given by Hansen in \cite{hansen}, is quoted.
For the definition one demands that $V$ is asymptotically flat in the following sense:
There exists a 3-manifold $\widetilde V$ and a conformal factor $\Omega$ satisfying
\begin{itemize}
\item[(i)]{$\widetilde V = V \cup \Lambda$, \; where $\Lambda$ is a single point}
\item[(ii)]{$\tilde h_{ab}=\Omega^2 h_{ab}$ is a smooth metric on $\widetilde V$}
\item[(iii)]{At $\Lambda$, $\Omega=0, \tilde D_a \Omega =0, \tilde D_a \tilde  
D_b \Omega = 2 \tilde h_{ab}$,}
\end{itemize}
where $\tilde D_a$ is the derivative operator associated with $\tilde h_{ab}$.
Note that we only demand asymptotic flatness on $V$. We do not demand that the spacetime 
 ``asymptotically tends to Minkowski spacetime''. Thus in this sense the
Taub-NUT solutions are asymptotically flat.

On $M$, and/or $V$ one defines the scalar potential 
\footnote{This definition differs from Hansen but is equivalent. 
See theorem 4 in \cite{simon}, or for the axisymmetric case \cite{backdahl2}.} 
$\phi$ as in \eqref{phidef}.
The multipole moments of $M$ are then defined on $\widetilde V$ as certain  
derivatives of the scalar
potential $\tilde \phi=\phi/\sqrt \Omega$ at $\Lambda$. More  
explicitly, following \cite{hansen}, let $\widetilde R_{ab}$ denote
the Ricci tensor of $\widetilde V$, and let $P=\tilde \phi$. Define the  
sequence $P, P_{a_1}, P_{a_1a_2}, \ldots$
of tensors recursively:
\begin{equation} \label{orgrec} P_{a_1 \ldots a_n}=C[
\tilde D_{a_1}P_{a_2 \ldots a_n}-
\tfrac{(n-1)(2n-3)}{2}\widetilde R_{a_1 a_2}P_{a_3 \ldots a_n}],
\end{equation}
where $C[\ \cdot \ ]$ stands for taking the totally symmetric and  
trace-free part. The multipole moments
of $M$ are then defined as the tensors $P_{a_1 \ldots a_n}$ at  
$\Lambda$. 

\section{Axisymmetry}
To simplify the situation we will henceforth only consider axisymmetric spacetimes.
In this situation there is a canonical way (see for instance \cite{wald}) 
of finding the Weyl-coordinates $(t,\varphi,R,Z)$,
where $(\frac{\partial}{\partial t})^a=\xi^a$ and $(\frac{\partial}{\partial \varphi})^a$ 
are the time-like and space-like Killing vector fields respectively.

In these coordinates the metric becomes
\begin{equation} \label{M}
ds^2=-\lambda(dt-Wd\varphi)^2+\lambda^{-1}(R^2d\varphi^2+e^{2\beta}(dR^2+dZ^2)),
\end{equation}
where the functions $\lambda$, $\beta$, and $W$ only depend on $R$ and $Z$.
We also observe that $\lambda$ is the norm of the time-like Killing vector field $\xi^a$.

The metric on $V$ becomes
\begin{equation}\label{orgmet}
h_{ab}\sim R^2d\phi^2+e^{2\beta}(dR^2+dZ^2) .
\end{equation}

In these coordinates, equation \eqref{ernst} reads
\begin{equation}\label{ernst2}
\phi_{RR}+\phi_{ZZ}+\tfrac{\phi_R}{R}=\frac{2\bar\phi}{\phi\bar\phi-1}\left ( \phi_R^2+\phi_Z^2\right ) .
\end{equation}
Introduce new coordinates $\tilde\rho=\frac{R}{R^2+Z^2}$, $\tilde z=\frac{Z}{R^2+Z^2}$ and rescale the potential
\begin{equation}\label{fdef}
f=\frac{\phi}{\sqrt{\tilde \rho^2+\tilde z^2}} .
\end{equation}

To be consistent with the notations in \cite{backdahl2}, we introduce different conformal factors.
An initial conformal transformation is formed by $\hat\Omega=(\tilde\rho^2+\tilde z^2)e^{-\beta}$.
We then get the rescaled metric of the desired form
\begin{equation} \label{omskmet}
\hat h_{ab}= \hat\Omega^2 h_{ab}\sim \tilde\rho^2 e^{-2\beta}d\varphi^2+d\tilde\rho^2+d\tilde z^2.
\end{equation}

The additional freedom in the choice of conformal factor can be described by letting $\Omega=e^{\kappa}\hat\Omega$,
where $\kappa(\Lambda)=0$. The vector $\kappa'(0)$ then corresponds to a translation. The higher order terms of $\kappa$
leaves the multipole moments unchanged and can therefore be freely specified. 
This freedom was used to simplify the recursive definition of the multipole moments in \cite{backdahl2}.
The metric corresponding to $\Omega$ is $\tilde h_{ab}=\Omega^2 h_{ab}=e^{2\kappa}\hat h_{ab}$
The potentials are rescaled as $\hat \phi=\phi/\sqrt {\hat\Omega}$, and $\tilde \phi=\phi/\sqrt \Omega$.
We note that $\tilde \phi=e^{\beta/2-\kappa/2}\frac{\phi}{\sqrt{\tilde\rho^2+\tilde z^2}}
=e^{\beta/2-\kappa/2} f$.

The axisymmetric condition also simplifies the description of the multipole moments.
We find \cite{geroch} that at $\Lambda$,  
$P_{a_1a_2\dots a_n}$ is proportional to $C[z_{a_1}z_{a_2}\dots z_{a_n}]$, i.e., 
we can define scalars $m_n$ via
\begin{equation}
P_{a_1a_2\dots a_n}(\Lambda)=m_n C[z_{a_1}z_{a_2}\dots z_{a_n}],
\end{equation}
where $z_a=(d\tilde z)_a$ is the direction along the symmetry axis.
Thus we may represent the multipole moments with $m_n$.

\noindent {\bf Remark.}
As well as different sign conventions some authors use the 
convention $M_n=\frac{1}{n!}\tilde z^{a_1}\dots\tilde z^{a_n}P_{a_1\dots a_n}(\Lambda)$.
This implies that $M_n=\pm \frac{2^nn!}{(2n)!}m_n$.

\section{Axisymmetric field equations}
For further studies we will need to reformulate the field equations. 
The new formulation is chosen such that the potential on a coordinate axis 
is the leading order part of the potential. This is useful because as we see 
later one can reconstruct the leading order part of the potential from the multipole moments.
The definition of the leading order part follows at the end of this section.

Henceforth we assume that $f$ is real-analytic in terms of 
$(\tilde x=\tilde\rho\cos\varphi, \tilde y=\tilde\rho\sin\varphi, \tilde z)$
and independent of $\varphi$. Thus we only search for solutions of this form.
These properties implies that $f$ is even in $\tilde\rho$, 
where the map $\tilde \rho \rightarrow -\tilde\rho$ is treated as 
$\varphi \rightarrow \varphi +\pi$.

Now define a complex variable $w=\frac{\tilde z +i\tilde\rho}2$.
Note that $f$ will not be analytic in $w$ but will have the form
\begin{equation}\label{gpredef}
f(w)=\sum_{n=0}^\infty\sum_{m=0}^\infty a_{n,m} w^n\bar w^m .
\end{equation}
The $\tilde \rho$-symmetry implies the symmetry $f(w)=f(\bar w)$, thus $a_{n,m}=a_{m,n}$.

Expressed in $w$ and $\bar w$ the equation \eqref{ernst2} reads
\begin{equation}\label{ernstomskpre}
f_{w\bar w}=\frac{f_w-f_{\bar w}}{2(w-\bar w)}-\frac{2\bar f (f+2w
  f_w)(f+2\bar w f_{\bar w})}{1-4w \bar w f\bar f}.
\end{equation}

Now we can define a function $g$ to be the function of two independent
real non-negative variables $\xi,\zeta$ defined by
\begin{equation}\label{gdef}
g(\xi,\zeta)=\sum_{n=0}^\infty\sum_{m=0}^\infty a_{n,m}\xi^n\zeta^m ,
\end{equation}
where $a_{n,m}$ is the same as in \eqref{gpredef}.
If \eqref{gpredef} converges for $|w| < r_0$ then 
\eqref{gdef} converges for $0\leq \xi,\zeta < r_0$ and vice versa.
This gives a bijection, mapping $f$ to $g$. Note that the
coefficients of $\bar f$ is $\bar a_{m,n}=\bar a_{n,m}$, thus
$\bar f$ is mapped to $\bar g$. The symmetry $f(w)=f(\bar w)$ implies the 
symmetry $g(\xi,\zeta)=g(\zeta,\xi)$.
Hence the equation \eqref{ernstomskpre} is equivalent to
\begin{equation}\label{ernstomsk}
g_{\xi\zeta}=\frac{g_\xi-g_{\zeta}}{2(\xi-\zeta)}-\frac{2\bar g (g+2\xi
  g_\xi)(g+2\zeta g_{\zeta})}{1-4\xi\zeta g\bar g}.
\end{equation}

\begin{definition} \label{ldef}
Suppose that $f$ is a real analytic function of 
$(\tilde x=\tilde\rho\cos\varphi, \tilde y=\tilde\rho\sin\varphi, \tilde z)$
independent of $\varphi$. We then define the leading order part of $f$ to be
$f_L(r)=g(r,0)$, where $g(\xi,\zeta)$ is the function formed from $f$ as above.
\end{definition}

Note that this is equivalent with the definition in \cite{backdahl2}. 
There we treated $f$ as a function of $(\tilde z,\tilde \rho)$, 
complexified $\tilde \rho$ via the power-series expression and defined
$f_L(r)=f(r,-ir)$.


\section{Specified multipole moments}
In this section we study the relation between the multipole moments and the leading order part 
of the potential $f$. 
How to compute the multipole moments from the leading order part of the potential and 
the metric component $\beta$, was studied in \cite{backdahl2}.
Therefore we cite the main theorem from \cite{backdahl2}.
\begin{theorem}\label{scalarfunc}
Suppose that $M$ is a stationary axisymmetric asymptotically flat spacetime, 
with $\tilde V$ being a conformal rescaling of the manifold of time-like Killing-trajectories,
where the metric $\hat h_{ab}$ on $\tilde V$ has the form \eqref{omskmet}. 
Let $\hat\phi$ be the conformally rescaled potential.
Furthermore, assume that $\hat\phi$ and $\beta$ are analytic 
in a neighbourhood of $\Lambda \in \tilde V$.
Choose $\kappa$ such that, for some constant $C$,
\begin{equation} \label{kappal}
\kappa_L(r)=-\ln(1-r\int_0^r{\frac{e^{2\beta_L(r')}-1}{{r'}^2}dr'}-r C)+\beta_L(r),
\end{equation}
thereby inducing the metric $\tilde h_{ab}=e^{2\kappa}\hat h_{ab}$.
Define $r(\rho)$ implicitly by $\rho(r)=re^{\kappa_L-\beta_L}$.
Put $y(\rho)=\tilde \phi_L(r(\rho))=e^{-\kappa_L(r(\rho))/2}\hat\phi_L(r(\rho))$. Then the multipole moments
$m_0,m_1,\dots$ of $M$ are given by $m_n=\frac{d^ny}{d\rho^n}(0)$.
\end{theorem}
Note that if we demand $f$ and $\beta$ to be analytic in $(\tilde x,\tilde y,\tilde z)$ and independent of $\varphi$,
the analyticity conditions of the theorem is fulfilled.

To be able to find a relation between the multipole moments and $f_L$ alone, we need to eliminate $\beta_L$ from 
theorem \ref{scalarfunc}.
With the choice of Weyl-coordinates this can be done as follows.
As in \cite{backdahl2} we let 
\begin{equation}
\eta^a=\left (\frac{\partial}{\partial \tilde z}\right )^a-i\left (\frac{\partial}{\partial \tilde \rho}\right)^a
=\left ( \frac{\partial}{\partial w}\right)^a
=(R+iZ)^2(\left(\frac{\partial}{\partial Z}\right)^a+i\left(\frac{\partial}{\partial R}\right)^a) .
\end{equation}
An explicit computation of the Ricci tensor for the 
metric \eqref{orgmet} implies
\begin{equation}
\eta^a\eta^bR_{ab}=\frac{2\bar w}{w(w-\bar  w)}\frac{\partial\beta}{\partial w}.
\end{equation}

From \eqref{riccieq} we get 
\begin{equation}
\eta^a\eta^b R_{ab}=\frac{2}{(1-\phi\bar\phi)^2}\frac{\partial\phi}{\partial w}\frac{\partial\bar\phi}{\partial w} .
\end{equation}
Combining this with \eqref{fdef} gives
\begin{equation}
\frac{\partial\beta}{\partial w}=\frac{w-\bar w}{(1-4 w\bar w f\bar f)^2}
(f+2 w \frac{\partial f}{\partial w})(\bar f+2 w \frac{\partial \bar f}{\partial w}) .
\end{equation}
The map $f$ to $g$  can also be done for $\beta$ to get a function $b(\xi,\zeta)$. For this function we get
\begin{equation}\label{betaprim}
\frac{\partial b}{\partial \xi}=\frac{\xi-\zeta}{(1-4 \xi\zeta g\bar g)^2}
(g+2 \xi \frac{\partial g}{\partial \xi})(\bar g+2 \xi \frac{\partial \bar g}{\partial \xi}) .
\end{equation}
This equation can then be used to reconstruct $b$ and hence $\beta$ from $g$.

The observation $\beta_L(r)=b(r,0)$ and $f_L(r)=g(r,0)$ implies
\begin{equation}\label{betaleq}
\beta_L'(r)=\Bigl(\frac{\partial b}{\partial \xi}\Bigr)_{\zeta=0,\xi=r}=r\Bigl|f_L(r)+2 r f_L'(r)\Bigr|^2 .
\end{equation}
This is easy to integrate to get $\beta_L$ and substitute this in theorem \ref{scalarfunc}. 
The constant of integration is chosen such that $\beta_L(0)=0$.
This must hold due to lemma 3 in \cite{backdahl2}, where it was proven that $\beta$ contains a factor $\tilde \rho^2$, 
thus $b$ contains a factor $(\xi-\zeta)^2$, i.e., $\beta_L(r)=\mathcal{O}(r^2)$.

Now we study the problem of finding the potential $f$ given the
multipole moments $m_n\in \mathbb{C}$. Form 
\begin{equation}\label{yrho}
y(\rho)=\sum_{n=0}^\infty \frac{m_n}{n!}\rho^n .
\end{equation}
We will assume that this power series converges in a 
neighbourhood of the origin.

To be able to express $f_L$ in terms of $y(\rho)$ we need the following relation between $\rho$ and $r$.
\begin{lemma} \label{rhorrel}
If $y$ is given by \eqref{yrho} and $\rho$, $\kappa$ are given implicitly by the relations in theorem \ref{scalarfunc}, then 
\begin{equation} \label{rhoekv}
0=-\frac{\rho}{r}+\rho \int_0^\rho \frac{1}{{\rho'}^2} \int_0^{\rho'} 2\rho'' 
\,|y(\rho'')+2\rho'' y'(\rho'')|^2\,d\rho''\, d\rho'+1+\kappa'(0)\rho,
\end{equation}
where the variables are real.
\end{lemma}
\begin{proof}
Theorem \ref{scalarfunc} gives the relation 
\begin{equation}\label{flyrel}
y(\rho)=e^{\tfrac{\beta_L(r(\rho))-\kappa_L(r(\rho))}{2}}f_L(r(\rho)) =\sqrt{\tfrac{r(\rho)}{\rho}}f_L(r(\rho)) .
\end{equation}
Combining this with \eqref{betaleq}, we have
\begin{equation}\label{betaprimy}
\beta_L'=\frac{r^2\rho_r^2}{\rho}\left | y+2\rho y_\rho \right |^2 .
\end{equation}
Together the definitions of $\rho$ and $\kappa_L$ gives
\begin{equation*}
\frac{r}{\rho}=e^{\beta_L-\kappa_L}=1-r\int_0^r{\frac{e^{2\beta_L(r')}-1}{{r'}^2}dr'}-r C,
\end{equation*}
dividing by $r$ and differentiating with respect to $r$ gives
\begin{equation}\label{betarhorel}
e^{2 \beta_L}=r^2\rho_r/\rho^2
\end{equation}
and a further differentiation with respect to $\rho$ gives the useful relation
\begin{equation} \label{beta}
- \frac{d}{d\rho}(e^{-2\beta_L(r)})=\frac{2\beta_L'(r)}{\rho_r}e^{2\beta_L(r)}=\frac{2\rho^2\beta_L'(r)}{r^2\rho_r^2} .
\end{equation}

Using this in \eqref{betaprimy}, we get
\begin{equation}
2 \rho |y+2\rho y_\rho|^2
=- \frac{d}{d\rho}(e^{-2\beta_L(r)}) .
\end{equation}
So that
\begin{equation}
e^{-2\beta_L(r)}=-\int_0^\rho 2 \rho |y+2\rho y_\rho|^2\,d\rho+C_1.
\end{equation} 
As proven in \cite{backdahl2}, $\beta_L(0)=0$, hence the limit $r \to 0$ gives $C_1=1$. 
Substituting \eqref{betarhorel} and dividing by $\rho^2$ gives
\begin{equation} \label{innanint}
\frac{1}{r^2 \rho_r}=-\frac{d}{d \rho} (\frac{1}{r})=
-\frac{1}{\rho^2} \int_0^\rho 2 \rho' |y(\rho')+2\rho' y'(\rho')|^2\,d\rho'+\frac{1}{\rho^2}.
\end{equation}
A further integration of \eqref{innanint} followed by a multiplication with $\rho$ yields
\begin{equation}
0=-\frac{\rho}{r}+\rho \int_0^\rho \frac{1}{{\rho'}^2} \int_0^{\rho'} 
2\rho'' \, |y(\rho'')+2\rho'' y'(\rho'')|^2\,d\rho'' \, d\rho'+1+C_2\rho.
\end{equation}
As in \cite{backdahl1} we find that $C_2=\kappa'_L(0)$ which is usually, but not always, put to 0. This gives lemma \ref{rhorrel}.
\end{proof}

Using this lemma it is not difficult to prove that $y(\rho)$ and thus
$m_n$ specifies $f_L$. In the next section we will see that the field
equations can be solved with the condition $g(r,0)=f_L(r)$.
\begin{theorem}\label{bmultipolethm}
Given a set of multipole moments $m_n$ such that \eqref{yrho} has positive radius of convergence, the corresponding
$f_L$ is uniquely specified and real-analytic in a neighbourhood of the origin.
\end{theorem}
\begin{proof}
The proof is essentially the same as in \cite{backdahl1}. From \eqref{rhoekv} we find that
\begin{equation}
r(\rho)=\rho/(1+\rho\mathfrak{g}(\rho)),
\end{equation}
where
\begin{equation}
\mathfrak{g}(\rho)=\int_0^\rho \frac{1}{{\rho'}^2} \int_0^{\rho'} 2\rho'' 
\,|y(\rho'')+2\rho'' y'(\rho'')|^2\,d\rho''\, d\rho'+1+\kappa'(0).
\end{equation}
is real-analytic in a neighbourhood of $\rho=0$. From $r(0)=0$, $r'(0)\neq 0$, 
the inverse function theorem for analytic functions gives that $\rho=\rho(r)$ is 
also a real-analytic function, with $\rho(0)=0$, in some neighbourhood of $r=0$.
The equation \eqref{flyrel} can be written
\begin{equation}
f_L(r)=\sqrt{\frac{\rho}{r}}y(\rho)=\sqrt{1+\rho(r)\mathfrak{g}(\rho(r))}y(\rho(r)) .
\end{equation}
Thus $f_L(r)$ is a complex valued real-analytic function in a neighbourhood of the origin.
\end{proof}

\subsection{Solving the field equations}
We can now turn our attention to the problem of extending $f_L$ to a solution $f$ of the field equations.
\begin{lemma}\label{formalsol}
Let $f_L(\xi)$ be analytic in a neighbourhood of the origin. 
Then there is a formal power series solution $g(\xi,\zeta)$ to \eqref{ernstomsk} such that $g(\xi,0)=f_L(\xi)$.
\end{lemma}
\begin{proof}
In a neighbourhood of the origin we can formally expand the non-linear term of \eqref{ernstomsk}.
Thus, define $c_{n,m}$ via
\begin{equation}\label{olin}
\sum_{n=0}^\infty\sum_{m=0}^\infty c_{n,m}\xi^n\zeta^m 
=\frac{2\bar g (g+2\xi g_\xi)(g+2\zeta g_\zeta)}{1-4\xi\zeta g\bar g}.
\end{equation}
Due to the symmetry in $\xi$ and $\zeta$, we have $c_{n,m}=c_{m,n}$.

Let $\mathcal{P}_n$ denote the set of polynomials in $a_{k,l}$ and $\bar a_{k,l}$ with 
non-negative coefficients, where $k+l \leq n$.
We show that $c_{n,m}=c_{n,m}(a_{.,.},\bar a_{.,.})\in \mathcal{P}_{n+m}$. 
The $\xi^n\zeta^m$-coefficient of $(g+2\xi g_\xi)(g+2\zeta g_\zeta)$ is easily found to be
\begin{equation}
\sum_{k=0}^n\sum_{l=0}^m a_{n-k,l}a_{m-l,k}(1+2(n-k))(1+2(m-l)) .
\end{equation}
Thus it is a polynomial in $a_{k,l}$ with non-negative coefficients, where $k+l\leq n+m$.
The $\xi^n\zeta^m$-coefficient of $(g\bar g)^p$, where $p\in \mathbb{N}$ is easily seen to be in $\mathcal{P}_{n+m}$.
Therefore, the $\xi^n\zeta^m$-coefficient of 
\begin{equation}
\frac{1}{1-4\xi\zeta g\bar g}=\sum_{p=0}^\infty (4\xi\zeta g\bar g)^p
\end{equation}
is also in $\mathcal{P}_{n+m}$.
Consider the set of power series in $\xi$ and $\zeta$, with the $\xi^n\zeta^m$-coefficient in $\mathcal{P}_{n+m}$.
This set is closed under products. Thus $c_{n,m} \in \mathcal{P}_{n+m}$.

The equation \eqref{ernstomsk} can be rewritten as the system 
\begin{equation}
(k+1)(n+\tfrac{3}{2})a_{n+1,k+1}=\sum_{m=0}^k (1-m+\frac{k+n}{2})a_{n+k+2-m,m}-c_{n,k}
\end{equation}
where $0\leq k\leq n$.

This then can be reformulated as
\begin{equation}\label{rec1}
a_{n-k,k}=\frac{(2k-1)(n-k+1)}{k(2n-2k+1)}a_{n-k+1,k-1}+2\frac{c_{n-k,k-2}-c_{n-k-1,k-1}}{k(2n-2k+1)}
\end{equation}
and simplified to
\begin{equation}\label{rec3}
\begin{split}
a_{n-k,k}&=\frac{(2k-1)!!(2n-2k-1)!!}{k!(n-k)!}\left ( \frac{n!a_{n,0}}{(2n-1)!!}
-\frac{2(k-1)!(n-k)!c_{n-k-1,k-1}}{(2k-1)!!(2n-2k+1)!!} \right . \\
& \left . 
-2 \sum_{i=1}^{k-1}\frac{(i-1)!(n-i-1)!(n-2i)}{(2i+1)!!(2n-2i+1)!!}c_{n-i-1,i-1}
\right ) ,
\end{split}
\end{equation}
where $1\leq k\leq \lfloor \tfrac{n}{2} \rfloor$.
The data $g(\xi,0)$ gives $a_{n,0}$ and together with \eqref{olin}, 
the recursion \eqref{rec3} gives a formal power series solution $g(\xi,\zeta)$.
\end{proof}

Now we want to establish convergence of the series produced by the previous lemma.
For this we will use the technique of majorizing series.
\begin{definition}
A series $\sum_{n=0}^\infty b_n$ with $b_n \geq 0$ is said to majorize a series
$\sum_{n=0}^\infty a_n$ with $a_n \geq 0$ if $b_n \geq a_n$ for all $n$.
\end{definition}

\begin{lemma}\label{convrec}
The formal solution from lemma \ref{formalsol} converges in a neighbourhood of the origin.
\end{lemma}
\begin{proof}
Now let $M,R > 0$ be such that $\frac{M}{\sqrt{1-\tfrac{\xi}{R}}}$ majorizes 
$\sum_{n=0}^\infty |a_{n,0}|\xi^n$.
This can always be done. For instance choose $R$ such that the series is bounded on $0\leq \xi \leq 2R$, 
and let $M$ be the maximal value of the series on that interval.

The triangle inequality on the recursive step \eqref{rec3}, together with the non-negativity 
of the coefficients of $c$, implies
\begin{equation}\label{majreca}
\begin{split}
|a_{n-k,k}|& \leq \frac{(2k-1)!!(2n-2k-1)!!}{k!(n-k)!}\left ( \frac{n!|a_{n,0}|}{(2n-1)!!}
+\frac{2(k-1)!(n-k)!|c_{n-k-1,k-1}|}{(2k-1)!!(2n-2k+1)!!} \right . \\
& \left . 
+2 \sum_{i=1}^{k-1}\frac{(i-1)!(n-i-1)!(n-2i)}{(2i+1)!!(2n-2i+1)!!}|c_{n-i-1,i-1}|
\right )\\
&\leq\frac{(2k-1)!!(2n-2k-1)!!}{k!(n-k)!}\left ( \frac{n!|a_{n,0}|}{(2n-1)!!} \right . \\
&+\frac{2(k-1)!(n-k)!c_{n-k-1,k-1}(|a_{.,.}|,|a_{.,.}|)}{(2k-1)!!(2n-2k+1)!!}\\
& \left . 
+2 \sum_{i=1}^{k-1}\frac{(i-1)!(n-i-1)!(n-2i)}{(2i+1)!!(2n-2i+1)!!}c_{n-i-1,i-1}(|a_{.,.}|,|a_{.,.}|)
\right ) .
\end{split}
\end{equation}
Now we want to construct a series majorizing $\sum_{n=0}^\infty\sum_{m=0}^\infty |a_{n,m}|\xi^n\zeta^m$.
Let $b_{n,k}\geq 0 $ satisfy
\begin{equation}\label{majrec}
\begin{split}
b_{n-k,k}& = \frac{(2k-1)!!(2n-2k-1)!!}{k!(n-k)!}\left ( \frac{n!}{(2n-1)!!}b_{n,0}\right . \\
&+\frac{2(k-1)!(n-k)!c_{n-k-1,k-1}(b_{.,.},b_{.,.})}{(2k-1)!!(2n-2k+1)!!}\\
& \left . 
+2 \sum_{i=1}^{k-1}\frac{(i-1)!(n-i-1)!(n-2i)}{(2i+1)!!(2n-2i+1)!!}c_{n-i-1,i-1}(b_{.,.},b_{.,.})
\right )
\end{split}
\end{equation}
and form 
\begin{equation}
h(\xi,\zeta)=\sum_{n=0}^\infty(b_{n,n}\xi^n\zeta^n+\sum_{m=0}^{n-1}b_{n,m}(\xi^n\zeta^m+\xi^m\zeta^n)) .
\end{equation}
By comparing with lemma \ref{formalsol}, we see that 
recursion \eqref{majrec} gives the coefficients of the analytic solutions of the equation
\begin{equation}\label{ernstmaj}
h_{\xi\zeta}=\frac{h_\xi-h_\zeta}{2(\xi-\zeta)}+\frac{2h (h+2\xi
  h_\xi)(h+2\zeta h_\zeta)}{1-4\xi\zeta h^2}.
\end{equation}
This equation can be simplified to a linear equation by a non-linear transformation. 
To find this transformation we substitute
\begin{equation}
h=\frac{{\mathfrak f}(\sqrt{\xi\zeta}\alpha(\xi,\zeta))}{\sqrt{\xi\zeta}} ,
\end{equation}
in \eqref{ernstmaj} and force the $\alpha_\xi\alpha_\zeta$-term to vanish.
The resulting differential equation for $\mathfrak f$ can then be solved to obtain the transformation
\begin{equation}
h(\xi,\zeta)=\frac{\sin(\tfrac{1}{3}\arcsin(3\sqrt{\xi\zeta}\alpha(\xi,\zeta)))}{\sqrt{\xi\zeta}} .
\end{equation}
Substituting this into \eqref{ernstmaj}, we get the linear equation
\begin{equation}
\alpha_{\xi\zeta}=\frac{\alpha_\xi-\alpha_\zeta}{2(\xi-\zeta)} .
\end{equation}
If we let $\alpha(\xi,0)=h(\xi,0)=\frac{M}{\sqrt{1-\tfrac{\xi}{R}}}$, we get the solution
$\alpha(\xi,\zeta)=\frac{M}{\sqrt{1-\tfrac{\xi}{R}}\sqrt{1-\tfrac{\zeta}{R}}}$, thus
\begin{equation}
h(\xi,\zeta)=\frac{\sin(\tfrac{1}{3}\arcsin(
\frac{3M\sqrt{\xi\zeta}}{\sqrt{1-\tfrac{\xi}{R}}\sqrt{1-\tfrac{\zeta}{R}}}))}{\sqrt{\xi\zeta}}.
\end{equation}
Note that this is real-analytic in $\xi$ and $\zeta$, with non-negative coefficients, and have 
the symmetry $h(\xi,\zeta)=h(\zeta,\xi)$. Furthermore it is bounded in 
the region $0\leq \xi,\zeta\leq \frac{R}{1+3MR}$.

The choice of $R,M$ gives $|a_{n,0}|\leq b_{n,0}$.
We will now prove that $|a_{n-k,k}|\leq b_{n-k,k}$ for all $0 \leq k \leq \lfloor \tfrac{n}{2} \rfloor$ 
and all $n\in \mathbb{N}$ by induction over $n$.
The base step ($n=0,1$) is trivial.
The induction hypothesis is that it holds for $n-2$. Together with the inequalities 
$|a_{n,0}|\leq b_{n,0}$ and \eqref{majreca} we have
\begin{equation}\label{majrecb}
\begin{split}
|a_{n-k,k}|& \leq\frac{(2k-1)!!(2n-2k-1)!!}{k!(n-k)!}\left ( \frac{n!}{(2n-1)!!}b_{n,0}\right . \\
&+\frac{2(k-1)!(n-k)!c_{n-k-1,k-1}(b_{.,.},b_{.,.})}{(2k-1)!!(2n-2k+1)!!} \\
& \left . 
+2 \sum_{i=1}^{k-1}\frac{(i-1)!(n-i-1)!(n-2i)}{(2i+1)!!(2n-2i+1)!!}c_{n-i-1,i-1}(b_{.,.},b_{.,.})
\right )\\
&=b_{n-k,k},
\end{split}
\end{equation}
where the last equality is \eqref{majrec}. 
Hence $|a_{n,k}|\leq b_{n,k}$ for arbitrary $n,k \in \mathbb{N}$ with $k \leq n$.
Thus the series expansion for $g$ converges whenever $h$ does.
\end{proof}
Note that by lemma \ref{formalsol} we can compute a number of terms of the power series $g$.
The modulus of the remaining terms can then be estimated by the methods in lemma \ref{convrec}.
However for the proof of existence of solutions we only need the convergence.

Now we can map $g(\xi,\zeta)$ back to $f(w)$ and change back to the variables
$(\tilde z, \tilde \rho)$ via $w=\tfrac{\tilde z+ i\tilde \rho}{2}$. 
The potential will still be analytic and even in $\tilde \rho$.
Therefore we have analyticity in $\tilde z$ and $\tilde \rho^2=\tilde x^2+\tilde y^2$. 
Thus, together with theorem \ref{bmultipolethm} we have
\begin{theorem}
For every set of multipole moments $m_n$ such that the power series \eqref{yrho} has 
positive radius of convergence, 
there is a unique solution $f$ to the field equations, in the class of analytic functions of
$(\tilde x,\tilde y,\tilde z)$.
\end{theorem}

Now we have the potential $f$, we easily find $\lambda$ and $\omega$. It can be shown that the metric 
function $\beta$ can always be obtained by integrating \eqref{betaprim}. The constant of integration can always be
chosen such that $b(\xi,\zeta)=b(\zeta,\xi)$ and that $\frac{b(\xi,\zeta)}{(\xi-\zeta)^2}$ is analytic in $\xi$ and $\zeta$.
With this choice the corresponding $\beta$ will be analytic in $(\tilde x, \tilde y, \tilde z)$, and tend to zero 
fast enough for $V$ to be asymptotically flat. Hence, the solution has all properties demanded in the theorems.
Therefore the multipole moments can be freely specified as long as the power series \eqref{yrho} has 
positive radius of convergence.

The metric function $W$ can also be obtained by an integration. If we don't demand $m_0$ to be real $W$ will not 
asymptotically tend to zero. In this case we get a non-zero NUT-parameter. 
\section{Conclusions and discussion}
We can conclude that if the multipole moments are chosen such that the power series \eqref{yrho} has positive 
radius of convergence, the corresponding potential function $f$ exist and is analytic in the rescaled 
spatial variables. Furthermore the corresponding metric functions for the spacetime can be found. 
Asymptotic flatness of the manifold of time-like Killing trajectories is guaranteed, but the spacetime 
will in general not be ``asymptotically Minkowski''.

The methods presented here can also be used to construct axisymmetric spacetimes with specified multipole
moments. The leading order part of the potential can be found by theorem \ref{bmultipolethm}. However this
involves a function implicitly defined, which can make explicit computations difficult.
The main difficulty is to solve the recursion in lemma \ref{formalsol}.
The recursion can however be used to get approximations to arbitrary order. 
The methods in lemma \ref{convrec} can then be used to get estimates of the solution or the rest terms.

For the future it is natural to try to extend the results to the general stationary case, 
without the axisymmetry condition. The field equations will then be much more complicated,
therefore this task will be difficult.
Another possibility is to extend the results of this paper to the electro-vacuum case.

\section*{Acknowledgements}
I am grateful to Magnus Herberthson for a good cooperation in our
previous works as well as support when I wrote this paper.
Thanks to Bengt Ove Turesson for many ideas that turned out to be
fruitful in this paper as well as previous works.

\end{document}